\documentclass[prb, reprint, superscriptaddress, longbibliography]{revtex4-2}
\usepackage{amsmath}
\usepackage{amssymb}
\usepackage{graphicx}
\usepackage[caption=false, position=top, singlelinecheck=off, justification=raggedright]{subfig}
\usepackage{multirow}
\usepackage{color}
\usepackage[dvipsnames]{xcolor}
\usepackage{ulem}
\usepackage{pst-node}
\usepackage{appendix}
\usepackage[unicode]{hyperref}
\usepackage[version=3]{mhchem}
\usepackage{braket}
\hypersetup{
	unicode=true,         	
	colorlinks=true,      	
	linkcolor=blue,		   	
	citecolor=blue,       	
	urlcolor=blue		   	
}

\begin{document}


\title{First-order transition into a topological superfluid state in an atom-cavity system}


\author{Hannah Kleine-Pollmann}
\affiliation{Zentrum f\"ur Optische Quantentechnologien and Institut f\"ur Quantenphysik, Universit\"at Hamburg, 22761 Hamburg, Germany}
\author{Ludwig Mathey}
\affiliation{Zentrum f\"ur Optische Quantentechnologien and Institut f\"ur Quantenphysik, Universit\"at Hamburg, 22761 Hamburg, Germany}
\affiliation{The Hamburg Centre for Ultrafast Imaging, Luruper Chaussee 149, 22761 Hamburg, Germany}
\date{\today}


\begin{abstract}
We propose to combine Bose-Einstein condensation in higher Bloch bands and a driven-dissipative cavity-BEC system into a hybrid light-matter platform. Specifically, the condensate is trapped in a bipartite $s$–$p_x$–$p_y$ lattice, with a tunable energy offset. This enables a controlled population transfer from the $s$-orbital to the nearly degenerate $p_x$ and $p_y$ orbitals. The system forms a chiral ground state with $p_x \pm i p_y$ symmetry, with staggered orbital currents.
By increasing the transverse pump strength, we drive the system into the superradiant phase, resulting in a self-organized, density checkerboard, which rectifies the staggered chiral order into a topological superfluid state.
Using truncated Wigner simulations and complementary mean-field analysis, we determine the phase transition into this state as first order.
Our results show that higher-band condensates coupled to a cavity provide a promising platform for engineering non-trivial orbital order and topological superfluid phases in quantum optical many-body systems.
\end{abstract}
\maketitle


\section{Introduction}
\label{sec:introduction}

Since the first experimental realisation of a Bose-Einstein condensate (BEC) in dilute atomic gases~\cite{1995ScienceWiemannandCornell, 1995PRLKetterle}, ultracold atoms have become a toolbox for studying and understanding macroscopic quantum phenomena~\cite{2012NatureBloch}.
These systems allow precise control over dimensionality, interactions and external potential, enabling the research of quantum phase transitions~\cite{2013NatureMathey, 2011ScienceSengstock, 2014NatureEsslinger} and collective behaviour~\cite{RMP2008Zwerger} in regimes that are difficult to access in traditional condensed-matter systems~\cite{2005APJaksch, 2005NatureBloch, 2007APLewenstein}.
The ability to trap a BEC in optical lattices led to the experimental realisation of the Bose-Hubbard model~\cite{2002NatureBloch}.

Most experiments with ultracold atoms in an optical lattice~\cite{2006RevModPhysOberthaler,1992PRLGrynberg,1993PRLHansch} have focused on the lowest Bloch band. However, higher orbital ordering was identified as relevant for phenomena like high-$T_c$ superconductivity and metal-isolator transitions~\cite{1998RevModPhysYoshinori,2000ScienceNagaosa,2006RevModPhysXiao-Gang}. This motivated the proposal of loading atoms in higher Bloch bands~\cite{2005PRAGirvin,2006PRALiu} and their experimental realisation by stimulated Raman transition~\cite{2007PRLBloch}. 
Since then, constant efforts to simulate electronic properties by gauge fields and synthetic dimensions~\cite{2016NatureZoller,2019NatureOzawa} were made, using lattice modulation techniques~\cite{2011ScienceSengstock, 2014NatureEsslinger} and laser-induced tunneling~\cite{2014PRLBloch, 2013PRLKetterle, 2015NatureAidelsburger, 2015ScienceFallani, 2015ScienceSpielmann, 2011NatureLin}.
Further, loading atoms in higher orbitals can lead to band crossing and the emergence of topological non-trivial order~\cite{2012NatureSun} and unconventional bosonic phases such as metastable chiral condensates in the second Bloch band~\cite{2011NatureHemmerich, 2015PRLHemmerich, 2020PRRHemmerich} and a $p$-orbital chiral Bose liquid~\cite{2014NatureLiu}.
Moreover, superlattice geometries in a square lattice have enabled a $\pi$-flux bosonic lattice~\cite{2016PRLMoraisSmith, 2016PRLLiu}.

Optical cavities provide strong light-matter interactions~\cite{2013RMPEsslinger, 2021APRitsch} and therefore engineer long range correlation order~\cite{2004PRARitsch, 2007PRLDomokos} mediated by the cavity. In such cavity QED systems, the coupling between atomic motion and cavity mode gives rise to self-organization and collective light matter phases~\cite{1999PRABerman}. 
An example is the Dicke model~\cite{1954PRDicke}, which consists of $N$ two-level systems coupled to a single quantized light field mode and undergoes a superradiant phase transition associated with spontaneous $\mathbb{Z}_2$ symmetry breaking. 
Experimentally it is realised by a BEC placed in a standing wave optical cavity and pumped transversely with a standing wave laser~\cite{2010NatureBaumann}. 
Since then, ultracold atoms coupled to a cavity have enabled the observation of several many-body and collective phenomena~\cite{PRL2020SchleierSmith, 2018PRXLev, 2021PRLHemmerich, 2022ScienceKessler, 2024PRACosme} and the proposal of rotational sensors~\cite{skulte2023quantumrotationsensorrealtime}.
%
\begin{figure}[t!]
	\centering
	\includegraphics[scale=1]{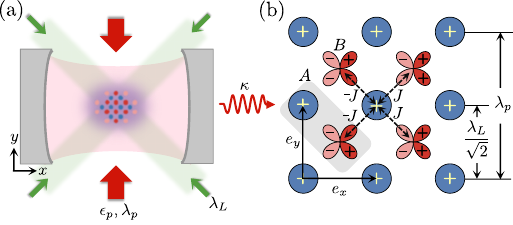}
    \caption{Sketch of the cavity-BEC system. (a) A BEC trapped in an optical lattice with wavelength $\lambda_\mathrm{L}$, placed in a high-finesse cavity and transversely pumped by a laser beam with wavelength $\lambda_p$ and pump strength $\epsilon_p$ along the $y$-direction. Photons leaking out of the cavity with the loss rate $\kappa$ along the $x$-direction. (b) Lattice geometry consisting of the two sublattices $A$ and $B$, with tunable relative potential depth $\Delta V$. Sublattice $A$ contains $s$-orbitals (circles), while sublattice $B$ contains $p_x$-$p_y$-orbitals (dumbbells). The orbital phase convention
    is indicated by $\pm$ signs, which determine the sign of the tunneling amplitude $J$ between neighbouring sites. The unit cell is indicated by the gray shaded rectangle. 
    } 
    \label{fig:fig1}
\end{figure}
%

In this paper, we propose to combine higher-orbital degrees of freedom with the enhanced light-matter interaction mediated by a cavity.
We demonstrate the realization of a self-organized topological superfluid state with a first-order transition, which is an unusual transition order in an ultracold atom context. 
Specifically, we consider an $s$-$p_x$-$p_y$ optical lattice, which breaks the $\mathbb{Z}_2$ symmetry through condensation into one of two degenerate chiral states.
We place this system inside a transversely pumped cavity, which independently breaks a $\mathbb{Z}_2$ symmetry upon entering the superradiant phase.
Both ingredients are experimentally motivated by experiments in Hamburg~\cite{2021KongkhambutPRL, 2011NatureHemmerich}.
We demonstrate that the simultaneous breaking of these $\mathbb{Z}_2$ symmetries in our three-dimensional system leads to a first-order transition and enables a self-organized chiral rectification.
The resulting state is associated with a non-zero Chern number~\cite{2016PRLLiu}. 

This paper is organized as follows. 
In Sec.~\ref{sec:model}, we introduce the model, lattice geometry and the simulation method.
In Sec.~\ref{sec:statepreparation}, we discuss the superradiant phase transition, the resulting symmetry broken ground states and determine the phase diagram. 
We identify the first-order nature of the phase transition.
In Sec.~\ref{sec:meanfieldcalculation} we perform a zeroth-order mean-field approximation and derive the corresponding energy functional.
Finally, in Sec.~\ref{sec:hysteresis} we present the hysteresis curve behaviour associated with the first-order transition. 


\section{Multi Orbital Dicke-Hubbard-Model}
\label{sec:model}

We consider a driven-dissipative atom-cavity system with a Bose-Einstein condensate (BEC) of $\ce{^{87}\mathrm{Rb}}$, trapped in a tunable optical lattice inside a high-finesse cavity and pumped along the $y$-direction, as schematically depicted in Fig.~\ref{fig:fig1}(a).
The tunability of the relative lattice depth between lattice site makes higher bands energetically accessible. This allows macroscopic occupation in these states.
We focus on the first excited band, which in a sufficiently deep lattice leads to the following extended Dicke-Hubbard Hamiltonian
\begin{align}
\label{eq:hamiltonian}
    \hat{H} =  \hat{H}_c +  \hat{H}_a+  \hat{H}_\mathrm{int}~,
\end{align}
where $\hat{H}_c$ describes the photon mode and its interaction with the atoms
\begin{align}
    \hat{H}_c/\hbar  &=-\Delta_c \hat{a}^\dagger\hat{a} +\Delta_0\hat{a}^\dagger\hat{a}\sum\limits_{\mathbf{r}} \Big\{\hat{n}_{p_x,\mathbf{r}} + \hat{n}_{p_y,\mathbf{r}} \Big\}
    \\&+g\big(\hat{a}^\dagger +\hat{a}\big)\sum\limits_{\mathbf{r}} (-1)^{\mathrm{r}_x+\mathrm{r}_y}\Big\{\hat{n}_{p_x,\mathbf{r}} + \hat{n}_{p_y,\mathbf{r}} \Big\}~.\nonumber
\end{align}
$\hat{H}_a$ describes the atoms without interactions:
\begin{align}
\hat{H}_a/\hbar=&\sum\limits_{\mathbf{r}}\Big\{ \tilde{V}_B\Big( \hat{n}_{p_x,\mathbf{r}} + \hat{n}_{p_y,\mathbf{r}} \Big)+ V_A \hat{n}_{s,\mathbf{r}} 
\\&-J \vphantom{\sum\limits_{\mathbf{r}}}\Big(\hat{b}^{\dagger}_{p_x,\mathbf{r}}\hat{b}_{s,\mathbf{r}+\mathbf{e}^\prime_y}-\hat{b}^{\dagger}_{p_x,\mathbf{r}}\hat{b}_{s,\mathbf{r}-\mathbf{e}^\prime_y}\nonumber
    \\&+\vphantom{\sum\limits_{\mathbf{r}}} \hat{b}^{\dagger}_{p_y,\mathbf{r}}\hat{b}_{s,\mathbf{r}+\mathbf{e}^\prime_x}-\hat{b}^{\dagger}_{p_y,\mathbf{r}}\hat{b}_{s,\mathbf{r}-\mathbf{e}^\prime_x}+\mathrm{h.c.}\Big)\Big\}\nonumber
\end{align}
and $H_\mathrm{int}$ describes the atom-atom interactions:
\begin{align}
H_\mathrm{int}/\hbar=   &\frac{U}{2} \sum\limits_{\mathbf{r}}\Big\{\hat{n}_{s,\mathbf{r}} \big( \hat{n}_{s,\mathbf{r}}-1\big) + \hat{n}_{p_x,\mathbf{r}}\hat{n}_{p_y,\mathbf{r}}
    \\& \quad+\vphantom{\sum\limits_{\mathbf{r}}}\frac{3}{4} \ \Big( \hat{n}_{p_x,\mathbf{r}} \big( \hat{n}_{p_x,\mathbf{r}}-1\big) + \hat{n}_{p_y,\mathbf{r}} \big( \hat{n}_{p_y,\mathbf{r}}-1\big)\Big)\nonumber
    \\& \quad+\vphantom{\sum\limits_{\mathbf{r}}}\frac{1}{4}\Big(\hat{b}^{\dagger}_{p_x,\mathbf{r}}\hat{b}^{\dagger}_{p_x,\mathbf{r}}\hat{b}_{p_y,\mathbf{r}}\hat{b}_{p_y,\mathbf{r}}+\mathrm{h.c.}\Big)\Big\}~,\nonumber
\end{align}
where $\mathbf{r}=(r_x,r_y)$ indexes the two dimensional lattice.
We emphasize that we consider two-dimensional lattices of one-dimensional tubes, extended along the z-direction.
The motion along the z-direction has been suppressed in the Hamiltonian for simplicity.
The bosonic operator $\hat{a}$ ($\hat{a}^\dagger$) annihilates (creates) a cavity photon and $\smash{\hat{b}^\dagger_{\smash{\{s,p_x,p_y\},\mathbf{r}}}}$ ($\smash{\hat{b}_{\{s,p_x,p_y\},\smash{\mathbf{r}}}}$) create (annihilates) an atom on site $\mathbf{r}$ in the respective orbital. 
The corresponding number operator is $\smash{\hat{n}_{\{s,p_x,p_y\},\smash{\mathbf{r}}}}$.
The pump-cavity detuning is $\Delta_\mathrm{c}<0$, the light-matter coupling we define as $g=\sqrt{\omega_\mathrm{rec}\epsilon_p|\Delta_0|}$ with $\Delta_0$ the light shift per intracavity photon. 
The pump laser intensity $\epsilon_p$ is measured in units of the recoil frequency $\omega_\mathrm{rec}$.
The on-site interaction strength is denoted by $U$ and the tunneling amplitude is given by $J$.

The lattice, schematically shown in Fig.~\ref{fig:fig1}(b), contains two inequivalent tube-shape lattice sites per unit cell, labeled as $A$ and $B$.
The unit vectors $\mathbf{e}_x$ and $\mathbf{e}_y$ define the lattice direction and $\mathbf{e}^\prime_y=(\mathbf{e}_x+\mathbf{e}_y)/2$ and $\mathbf{e}^\prime_x=(\mathbf{e}_x-\mathbf{e}_y)/2$ connect $A$ and $B$ sites. 
We assume $s$-orbitals on $A$-sites with potential $V_A$ and $p_x$ and $p_y$-orbitals on $B$-sites with potential $V_B$. 
We define tunneling between an $s$-orbital and the positive lobe of a $p$-orbital as $-|J|$, and tunneling into the negative lobe as $+|J|$, resulting in an alternating sign pattern of $|J|$ along the diagonal direction, see Fig.~\ref{fig:fig1}(b) for sign convention.

Note that the pump and the lattice laser intersect at an angle of $45^{\circ}$, with their wavelength chosen such that $\lambda_p=\lambda_L/\sqrt{2}$. 
This ensures that the periodicities of the pump and lattice potential are commensurate.
As a result, one sublattice, e.g. the $B$ sites, matches with the intensity maxima or minima of the optical lattice formed by pump laser and cavity, while the other sublattice, e.g. the $A$ sites, sits near the nodes.
This leads to a shifted on-site potential $\tilde{V}_B$ for the $B$-sites by the pump laser $\Delta_p=\mathrm{sign}(\Delta_0)\omega_\mathrm{rec}\epsilon_p$, i.e. we have $\tilde{V}_B = V_B + \Delta _p$.


\subsection{Experimental parameters and simulation method}
\label{subsec:experimentalparameter}
%
\begin{figure*}[t!]
    \centering
    \includegraphics[scale=1]{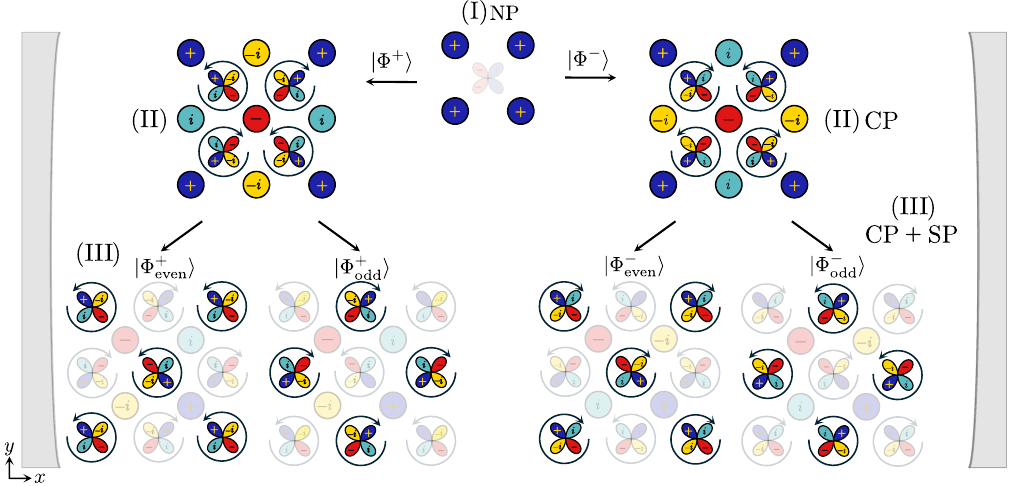}
    \caption{Symmetry-breaking sequence in the $s$–$p_x$–$p_y$ lattice coupled to an optical cavity (gray regions indicate cavity mirrors). Reduced orbital visibility reflects lower occupation.
(I) Normal Phase (NP): Atoms occupy $s$-orbitals on $A$-sites without phase coherence. Time-reversal and inversion symmetry are preserved.
(II) Chiral Phase (CP): tuning the relative sublattice potential depth $\Delta V$ leading atoms to condense into chiral states $p_x \pm i p_y$, breaking time-reversal symmetry and generating local orbital currents (indicated by circular arrows).
(III) Topological superfluid state, emerging due to chiral and superradiant ordering (CP + SP): Increasing pump strength $\epsilon_p$ induces a Dicke-type superradiant transition. The system condenses on even or odd $B$-sublattice sites, breaking lattice and time-reversal symmetry. The resulting states $\lvert \Phi^{\pm}_\mathrm{even} \rangle$ and $\lvert \Phi^{\pm}_\mathrm{odd} \rangle$ reflect this sublattice selectivity.}
    \label{fig:fig2}
\end{figure*}
%

We consider experimentally realistic parameters based on the experiments in Ref.~\cite{2021KongkhambutPRL, 2011NatureHemmerich}.
The atom number is set to $N_a=60 \times 10^3$, the recoil frequency of the pump laser to $\omega_\mathrm{rec}=2\pi \times 3.55$ kHz, the cavity photon loss rate to $\kappa = 2\pi \times 5.0$ kHz operating in the recoil-resolved regime.
For the optical lattice we use a laser with a wavelength of $\lambda_L=1064$ nm and for the retro reflected pump laser $\lambda_p=807$ nm, operating in the red-detuned regime compared to the relevant atomic transition.
The light shift per atom is $\Delta_0=-2 \pi \times 0.36$ Hz with an effective negative detuning $\Delta_\mathrm{eff}=\Delta_c-(1/2)N_a\Delta_0=-2 \pi \times 22.0$ kHz.
For the lattice parameter we assume $J \approx 2 \pi \times 202$ Hz for the tunneling rate and for the effective contact interaction $U \approx 2 \pi \times 71 $ Hz.
Note that the atoms are weakly confined in $z$-direction in a quasi 1D tube for which we account by artificially discretizing the $z$-dimension, see App.~\ref{app:sec-tubes}. 
For this we replace $U$ by the effective contact interaction $U_\mathrm{eff}$, which depends on the number of these effective layers per tube.

We simulate the dynamics in the semi-classical limit.
Therefore we employ the truncated Wigner approximation~\cite{2010APPolkovnikov} (TWA), which treats the quantum operators as $c$-numbers, leading to a set of coupled equations of motions given by,
\begin{align} 
        \frac{\partial}{\partial t} \hat{b}_{\mathbf{r},\nu}&=\frac{i}{\hbar} \frac{\partial H}{\partial \hat{b}^{*}_{\mathbf{r},\nu}}~,
    \\\frac{\partial}{\partial t} \hat{a}&=\frac{i}{\hbar} \frac{\partial \hat{H}}{\partial \hat{a}^*}-\kappa \hat{a}+\xi~,
\end{align}
where $\kappa$ is the photon loss rate via cavity and the corresponding stochastic noise term $\xi(t)$ with $\braket{\xi^*(t)\xi(t')}=\kappa \delta(t-t')$.
This method provides a semi-classical phase space description and is suitable for systems with a large number of atoms $N$. 
We initialize the cavity field by sampling from the Wigner distribution of a the vacuum state with $\braket{\alpha}=0$, with the noise corresponding to the quantum noise of an empty cavity.
For the atoms in the lattice, we sample the initial state from the Wigner distribution of a coherent state with a mean occupation $\braket{n_i}=N_a/M$, where $N_a$ is the total atom number and $M$ the number of lattice sites.


\section{State preparation and Superradiant Transition }
\label{sec:statepreparation}

We now discuss the state preparation of atoms in higher bands and recall the resulting ground state as realized and observed in Ref.~\cite{2011NatureHemmerich,2015PRLHemmerich}. 
After that we present the driving protocol for the transiton into the superradiant phase.
In Fig.~\ref{fig:fig2}, we sketch the different stages of our system in which we (I) initialise, (II) populate the $B$-sites and (III) transition into the superradiant phase (SR).
The cavity is indicated in grey in $x$-direction and the pump beam is in $y$-direction.  
\\
We initialize the system in the normal phase (NP) corresponding to a regime of on-site potentials with $V_A\ll V_B$ with an empty cavity field.
This amounts to all atoms initially occupying the lowest bands on the $A$ sites, corresponding to a state of independent quasi-condensates in the s-orbitals
The $B$ sites are not occupied, indicated by the low opacity in Fig.~\ref{fig:fig2}~(I).
\\
\begin{figure*}[t]
    \centering
    \includegraphics[scale=1]{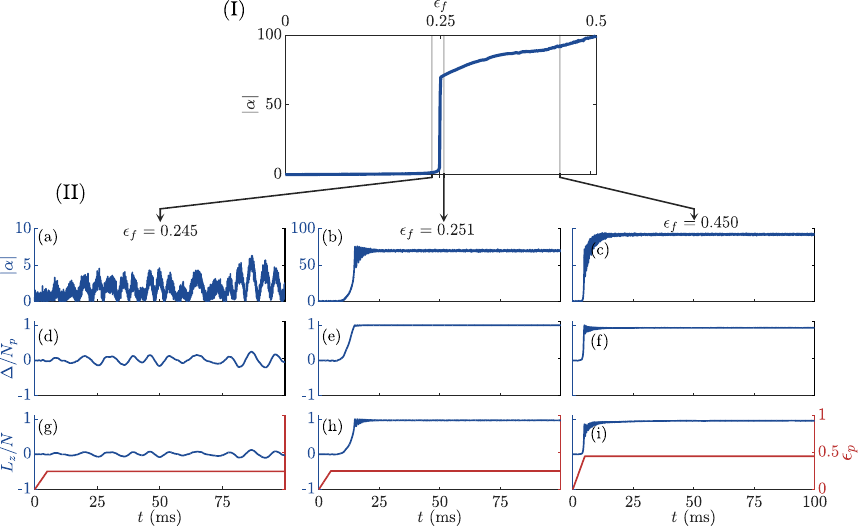}
    \caption{(I) Dynamical emergence of the topological  superfluid phase. We show the steady-state mean light field amplitude $|\alpha|$ as a function of the maximal pump strength $\epsilon_f$, in panel (I). The system evolves until it reaches a dynamical steady state. Each data point is the average over 100 trajectories using Truncated Wigner Approximation (TWA). (II) Individual TWA trajectories (blue) for three exemplary final pump strengths $\epsilon_f$ as a function of time $t$ in ms: (a,d,g) directly below the phase transition at $\epsilon_f=0.245$ , (b,e,h) directly above at $\epsilon_f=0.251$, and (c,f,i) far above at $\epsilon_f=0.450$. The pump beam protocol $\epsilon_p$ is indicated in red. We linearly increase $\epsilon_p$ within $5$ ms to its final values $\epsilon_f$ in each run. After reaching $\epsilon_f$, we hold it constant until the end of the simulation. Panel (a-c): We show the light field amplitude $|\alpha|$. Panel (d-f): We show the occupation imbalance between the two possible self-organized checkerboard patterns, normalised by $N_p$ the total number of particles in the $p_x$ and $p_y$-orbital. (g-i) We show the total angular momentum $L_z$ per particle in the $p$-orbitals in the lattice. }
    \label{fig:fig3}
\end{figure*}
%
Next, we linearly decrease the potential offset until both sites are energetically nearly degenerate and have transferred half of the occupation into the excited band. 
During this ramp, the cavity is strongly coupled to the vacuum mode. 
After that we use imaginary time propagation to converge into the lowest energy state. 
This chiral phase (CP) is characterized by chiral phase winding, clockwise or counter-clockwise on-site orbital flux, in the $p$-orbitals, see Fig.~\ref{fig:fig2}~(II).
The pair exchange term $\hat{b}^{\dagger}_{p_x,\mathbf{r}}\hat{b}^{\dagger}_{p_x,\mathbf{r}}\hat{b}_{p_y,\mathbf{r}}\hat{b}_{p_y,\mathbf{r}}$ induces a relative phase factor $\pm i$ leading to a $p_x \pm i p_y$ superposition state in staggered spatial order due to the staggered sign associated with the tunneling energy $J$, see App.~\ref{app:sec-groundstate} for details.
results in a spontaneously broken $\mathbb{Z}_2$ symmetry, such that the atoms condense in the $\ket{\Phi^+}$ or $\ket{\Phi^-}$ mode, see the left and right panel of Fig.~\ref{fig:fig2}~(II), respectively.
The two symmetry broken states can be distinguished by the chiral order parameter
\begin{align}
    \mathcal{C}_\mathrm{chir}=\frac{1}{N_p}\sum_\mathbf{r}(-1)^{r_x+r_y}\hat{L}_{z,,\mathbf{r}}~,
\end{align}
where $\hat{L}_{z,,\mathbf{r}}=-i\big(\hat{b}^{\dagger}_{p_x,\mathbf{r}}\hat{b}_{p_y,\mathbf{r}}-\hat{b}^{\dagger}_{p_y,\mathbf{r}}\hat{b}_{p_x,\mathbf{r}}\big)$ is the on-site angular momentum operator and takes values close to $ \pm 1$ for the $\ket{\Phi^\pm}$-mode.
The total angular momentum $\hat{L}^\mathrm{tot}_z$ vanishes.

With the system initialized in a chiral condensation mode, we linearly increase the pump intensity $\epsilon_p$ for 5~ms and keep constant.
The system realizes the chiral superradiant phase if $\epsilon_p$ exceeds a critical value $\epsilon_{p,c}$, see Fig.~\ref{fig:fig2}~(III).
A second $\mathbb{Z}_2$ symmetry is spontaneously broken due to the atoms locating on either the even or odd sites of the $B$-sublattice, consistent with the checkerboard lattice formed by the interference of pump and cavity fields.
This self-organization leads to a non zero total angular momentum $\hat{L}^\mathrm{tot}_z$ due to the population imbalance between neighbouring $B$-sites which have opposite local orbital flux.
We identify four distinct states $\ket{\Phi^\pm_\mathrm{even/odd}}$. 
The resulting state, which breaks a $\mathbb{Z}_2\otimes\mathbb{Z}_2$ symmetry, is a self-organized topological superfluid state. 
The superradiant order generates a checkerboard potential, which rectifies the staggered chirality, such that the total angular momentum acquires a non-zero value. 
The emergent state resembles the topological superfluid state put forth in Ref.~\cite{2016PRLLiu}, in which the chirality is rectified explicitly, by applying an external potential.

As we discuss in the following, the two ordering transition, each breaking a Z2 symmetry, synchronize, resulting in a single, first-order transition.


\subsection{Dynamical emergence of the topological superfluid state}
\label{sec:superradiance}

We now explore the dynamics of the system under constant driving and in the topological superfluid.
In our simulation, we initialize in the CP by randomly selecting one of the condensation modes $\ket{\Phi^\pm}$ then linearly increase $\epsilon_p$ to a final value $\epsilon_f$, that is kept constant after that, as described in the section before.
We determine the phase diagram with a $35\times35 \times 10$ $(x,y,z)$ lattice with stochastic noise in the intracavity light field.

A key observable is the occupation of the intracavity field $|\alpha|^2$ which distinguishes the normal, $|\alpha|^2=0$, and the superradiant $|\alpha|^2>0$ phase.
In addition, we are interested in the total angular momentum $L_z$ and the population imbalance $\Delta$ between the even and odd $B$-sublattice sites, which characterizes the state of the atomic sector.
We record these quantities when numerically solving the equations of motion.

In Fig.~\ref{fig:fig3}~(I), we show the steady state intracavity field amplitude $|\alpha|=\sqrt{\braket{\alpha}^2}$ as a function of the maximal pump strength $\epsilon_f$.
We time average the quantity after the system reaches a steady state, over 10 ms and taking $10^2$ trajectories.
Below a critical value $\epsilon_{c,p}$ we obtain an empty intracavity field as soon as $\epsilon_p$ exceed $\epsilon_{c,p}$, we obtain a finite intracavity field amplitude $|\alpha|$. 
We identify a step wise emerging $|\alpha|$ close to 0.25 which we relate to a first-order phase transition, which we explore in Sec.~\ref{sec:meanfieldcalculation}.  

In Fig.~\ref{fig:fig3}~(II), we present exemplary dynamics for different strengths $\epsilon_p$ of the pump laser with the driving protocol is depicted in red. 
We select trajectories initially in a $\ket{\Phi^+}$-state with positive chiral orientation and transition in a $\ket{\Phi^+_\mathrm{even}}$-state, corresponding to condensation on even lattice sites. 
We choose $\epsilon_p$ close to the phase boundary in the normal chiral phase and the topological superfluid phase, specifically $\epsilon_p=0.245$ and $\epsilon_p=0.251$, and far in the topological superfluid phase at $\epsilon_p=0.45$.

Driving at $\epsilon_f=0.245$ close to the SR transition, we observe the onset of fluctuations in the light field amplitude $|\alpha|$, as depicted in Fig.~\ref{fig:fig3}~(a). 
A similar fluctuating behaviour is displayed by the imbalance $\Delta$ and the total angular momentum $L_z^\mathrm{tot}$, as shown in Fig.~\ref{fig:fig3}~(d) and Fig.~\ref{fig:fig3}~(g). 
The magnitude of the fluctuations is influenced by the magnitude of the light field $|\alpha|$.
This reflects that the system is already influenced by the pump. However, the fluctuating behaviour leads to a vanishing time average of $|\alpha|$, in this regime, implying no order.

At $\epsilon_f=0.251$ above the SR phase boundary, we observe a persistent intracavity field amplitude $|\alpha|$, see Fig.~\ref{fig:fig3}~(b).
With a response delay of approximately 5 ms the cavity light field orders, indicated by $|\alpha|$ becoming finite with a transient oscillation which decays on a timescale of 25 ms, $|\alpha|$ remains finite. 
Simultaneously, $\Delta$ and $L_z^\mathrm{tot}$ converge to a positive value close to unity, indicating condensation on even sites and almost depopulated odd sites.
We further note that both exhibit the same transient behaviour as $|\alpha|$, see Fig.~\ref{fig:fig3}~(e,h).
The system undergoes spontaneous $\mathbb{Z}_2$ symmetry breaking into a topological superfluid state, as explained before.

In the regime, far above the SR phase boundary, at $\epsilon_f=0.450$ the system response is essentially instantaneous, see Fig.~\ref{fig:fig3}~(c,f,i). 
All three quantities $|\alpha|,\Delta$ and $L_z^\mathrm{tot}$ rise almost instantaneously and saturate at their final values but still show transient behaviour which decays within 10 ms.
Notably the final value of $\Delta$ and $L_z^\mathrm{tot}$ is slightly below the previous near threshold case.
We attribute this to a reduced phase coherence between sites.

We note that exemplary trajectories of the other symmetry-broken ground states ($\Phi^+_\mathrm{odd},\Phi^-_\mathrm{even},\Phi^+_\mathrm{odd}$) display similar behaviour.


\section{Mean field theory}
\label{sec:meanfieldcalculation}

\begin{figure}[t!]
	\centering
	\includegraphics[scale=1]{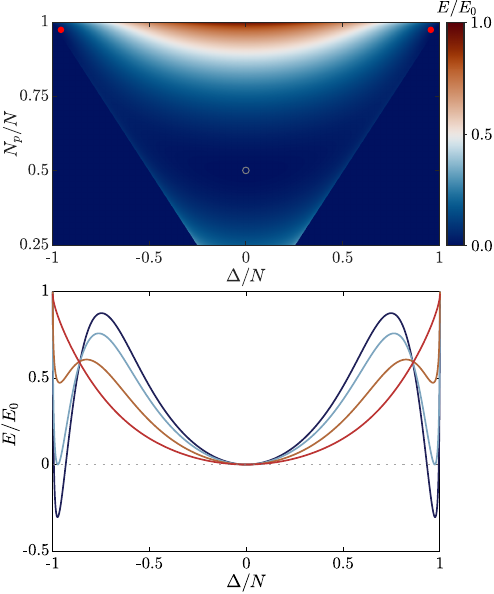}
    \caption{Mean‐field energy landscape $E/E_0$ as a function mean $p$‐orbital occupancy $\bar N_p$ and imbalance $\Delta$. For coupling $g<g_c$, the single minimum at $(\bar N_p=N/2,\Delta=0)$ marks the normal phase (grey circle). For $g>g_c$, two degenerate minima (red dots) at finite $\pm\Delta$ emerge, indicating the superradiant phase transiton.  $\bar N_p=N/2,\Delta=0$ remain a local minima (grey circle). Unphysical regions $|\Delta|>\bar N_p$ are shaded dark blue. (b) $\mathrm{min}_{\Bar{N}_p}E^\mathrm{MF}(\Delta)$ as a function of $\Delta$ for various $g$. The curves evolve from a parabola ($g = 0$, red) to a $\Phi^6$-type potential. At $g = g_c$, three degenerate minima appear. For $g > g_c$ (blue), two global minima emerge at finite $\Delta$ and a local minimum remains at $\Delta = 0$.
    } 
    \label{fig:fig4}
\end{figure}

We perform a zeroth-order mean-field (MF) approximation to support our statement regarding the first-order SR transition that we obtain numerically.
Therefore, all bosonic operators are replaced by their corresponding complex coherent-state amplitudes, thereby neglecting quantum fluctuations, i.e. $\braket{\hat{a}}=\alpha$, $\braket{\hat{b}_{\{s,p_x,p_y\},\mathbf{r}}}=\beta_{\{s,p_x,p_y\}}$ and $\braket{\hat{A}\hat{B}}= \braket{\hat{A}}\braket{\hat{B}}$ starting from Eq.~\ref{eq:hamiltonian}. 
We choose as the initial state one of the chiral $\mathbb{Z}_2$-symmetry-broken $\Phi^\pm$ states, so that each of the $B$-sites hosts a $\beta_\pm=\ket{p_x \pm i p_y}$ condensate with a fixed relative phase of $\pm \pi/2$, see App.~\ref{app-sec:meanfield} for details.
Without loss of generality, we set $V_A=0$, as only the relative detuning $\Delta V$ determines the relative occupation of sites $A$ and $B$.
We define $\beta_s=\sqrt{N_s/M_A}$ and $\beta_\pm=\sqrt{(N_p\pm \Delta)/M_B}$, where $N_s$ is the total particle number in the $s$-orbitals, $\bar{N}_p$ the total particle number in the $p$-orbitals and $M_B$ ($M_A$) denotes the total number of $B$-sites ($A$-sites).
We obtain the mean-field energy functional as
\begin{align}
    E^\mathrm{MF}&(\alpha,\Delta,\Bar{N}_p)
    =-\Delta_c\,\lvert \alpha\rvert^{2} +2g\Delta\alpha_r \\
    &
    -\tfrac{U}{2}(N-\Bar{N}_p)
    +(V_B+\Delta_0 \lvert \alpha\rvert^{2}-\tfrac{3U}{8})\Bar{N}_p\nonumber\\
    &
     -4J \sqrt{\tfrac{M_B}{2M_A}} \sqrt{N - \bar{N}_p} \Bigl( \sqrt{\bar{N}_p + \Delta} 
     + \sqrt{\bar{N}_p - \Delta} \Bigr)\nonumber \\
    &
    + \tfrac{U}{2M_A}(N-\Bar{N}_p)^2
    +\tfrac{U}{4M_B}(\Delta^2+\Bar{N}_p^2)~,\nonumber
\end{align}
Here, $\alpha_r=\mathcal{R}(\alpha)$ denotes the real part of the intracavity field.
We eliminate the cavity field $\alpha$ by minimizing $ \partial E^\mathrm{MF}/\partial\alpha$ which yields
\begin{equation}
    \alpha(\Delta,\Bar{N}_p)= \frac{g \Delta}{-\Delta_c+\Delta_0\Bar{N}_p}~.
\end{equation}
This substitution simplifies the energy functional to the remaining free parameters $\Delta,\Bar{N}_p$.
\\
In Fig.~\ref{fig:fig4}~(a), we shows the mean-field energy $ E^\mathrm{MF}(\Delta,\Bar{N}_p)$ as a function of mean $B$-site occupation $\Bar{N}_p$ and the imbalance $\Delta$.
We choose $g$ slightly above the critical threshold $g_c$, i.e. close to the SR phase transition. 
We identify two symmetric degenerate energy minima at $\pm \Delta \neq 0$ (red dots), indicating the $\mathbb{Z}_2$ symmetry breaking.
For references at $g=0$ the global energetic minimum sits at $\Delta=0$ and $N_p=N/2$ indicated by the grey circle.
We observe that crossing $g_c$ causes a discontinuous jump of both $\Delta$ and $\Bar{N}_p$.
The former global minimum at $\Delta=0$ and $N_p=N/2$ (grey circle, at $g=0$) becomes a local minimum for $g>g_c$.
Regions with $|\Delta|>\Bar{N}_p$ are unphysical (dark blue shaded).

In Fig.~\ref{fig:fig4}~(b), we show the reduced energy functional $\mathrm{min}_{\Bar{N}_p}E^\mathrm{MF}(\Delta)$ for several values of $g$. 
We minimise $E^\mathrm{MF}$ with respect to $\Bar{N}_p$.
At $g=0$ (dark red) we observe an approximately parabolic behaviour with one global minimum.
Increasing $g$ deforms the potential into a $\Phi^6$-type function.
We visualise the occurrence of two additional local minima by fine-tuning to the value $g=87.86097519899$ (light red), which is in the normal phase without a finite cavity field. 
The three minima become degenerate at $g_c= 89.86097519899065$ (light blue), which is directly at the phase transition. We note that the necessity of fine-tuning close to the phase boundary, is indicative of the rapidity of the transition.
For larger $g$, e.g.  $g=90.3$ (dark blue), the two symmetric minima at $\pm \Delta$ become global while $\Delta=0$ remains a local minimum. 
The transition from a $\Phi^6$-type potential is consistent with our numerical observation of a first-order phase transition~\cite{1937Landau}.
We use the same parameters as in our numerical simulations, with the exception of $V_B$. This was adjusted to match the occupation per site at $g = 0$, ensuring comparability between mean-field and numerical results.


\section{Hysteretic dynamics}
\label{sec:hysteresis}

\begin{figure}[t!]
	\centering
	\includegraphics[scale=1]{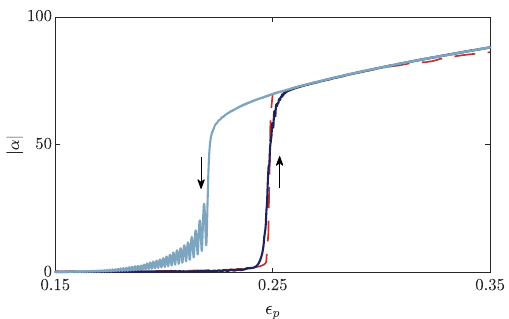}
    \caption{Hysteresis of the intracavity field amplitude $|\alpha|$ as a function of the pump strength $\epsilon_p$. The pump strength is linearly increased from $\epsilon_p=0$ to $0.351$ over $300\,\mathrm{ms}$ (up-ramp) and, after a hold time of $100\,\mathrm{ms}$, decreased back to zero over the same duration (down-ramp).} 
    \label{fig:fig5}
\end{figure}

We probe the system for for hysteretic dynamics by adiabatically varying the pump strength $\epsilon_p$ while simultaneously tracking the intracavity field amplitude $|\alpha|$.
Starting at $\epsilon_p=0$, we linearly increase $\epsilon_p$ to 0.351 over a duration of 300 ms followed by a waiting time of 100 ms. 
Subsequently, we adiabatically decrease the pump strength $\epsilon_p$ in 300 ms to zero.
The direction of the ramp is indicated by arrows in Fig.~\ref{fig:fig5} and the steady-state result is shown in red for reference.
We observe a significant hysteresis curve.
The transition from the normal phase to the chiral superradiant phase, shown in dark blue, occurs at a higher critical pump strength than the reverse transition, light blue, during the down-ramp.
The system follows distinct branches depending on the direction of parameter variation. This behaviour demonstrates that the state of the system depends not only on the current control parameter $\epsilon_p$ but also on its history.
While for the standard second-order  Dicke transition, a power-law closing of the hysteresis was demonstrated in Ref.~\cite{2015PNASKlinder}, consistent with Kibble-Zurek scaling of a second order transition, the persistence of the hysteresis to these long times is consistent hysteretic dynamics of a first-order transition.
%

\section{Conclusion}

In conclusion, we have put forth a atom-cavity system that displays a first-order transition into a self-organized topological superfluid state.
We propose to combine the cold-atom platforms established in Ref.~\cite{2011NatureHemmerich} and Ref.~\cite{2021KongkhambutPRL}.  
Specifically, the atoms are held in an optical lattice, and promoted to the first excited band, composed of s- and p-orbitals in a checkerboard pattern, via an 'elevator trick' of quenching the lattice structure.
The atoms form a chiral condensate in this band, in which the orbital angular momentum on each p-site acquires a non-zero value, in a staggered pattern. 
Additionally the atoms are coupled dispersively to an optical cavity and pumped from the transverse direction, resulting in a superradiant transition. 
In this transition the atoms spontaneously form a checkerboard density pattern, in addition to the photon mode of the cavity being populated. 
We propose to align the optical lattice to the cavity mode such that the chiral pattern of non-zero angular momenta is rectified by the formation of the density pattern in the superradiant transition. 
With this, the resulting state is a topological superfluid, that self-organizes in the pumped cavity. 
As we have shown in this paper, the two phase transitions synchronize into a single phase transition that is of first order. One way to detect this first-order character, is to perform a hysteretic measurement. 
With this proposal, we have put forth a hybrid cold atom platform that supports the investigation of physical phenomena such as chiral-to-topological transitions, light-matter dynamics involving higher bands, and first-order  criticality.


\begin{acknowledgments}
We thank Andreas Hemmerich and Max Hachmann for the fruitful discussions.
This work is funded by the University of Hamburg.
We acknowledge funding by the Cluster of Excellence
“Advanced Imaging of Matter” (EXC 2056) Project
No.390715994 and ERDF of the European
Union and by ’Fonds of the Hamburg Ministry of Sci-
ence, Research, Equalities and Districts (BWFGB)’.
\end{acknowledgments}


\bibliography{references}


\appendix


\section{Atom-Cavity Model}
\label{app:sec-model}

We start our theoretical description with a many body Hamiltonian, including the cavity field, atomic degrees of freedom, atom-light coupling and short-range atom-atom interactions.
\begin{align}
    \hat{H}&/\hbar= - \Delta_\mathrm{c} \hat{a}^\dagger\hat{a}
        +\int dx dy \ \hat{\Psi}^\dagger(\mathbf{r}) \Bigl[ -\frac{\hbar^2}{2m}\nabla^2+V(\mathbf{r})\nonumber
        \\&
        +\omega_\mathrm{rec}\epsilon_p\cos^2(k_py)
        + \Delta_0\hat{a}^\dagger\hat{a}\cos^2(k_px)\nonumber
        \\&
        +\sqrt{\epsilon_p\omega_\mathrm{rec}|\Delta_0|}(\hat{a}^\dagger+\hat{a})\cos(k_px)\cos(k_py)\Bigr]\hat{\Psi}(\mathbf{r})\nonumber
        \\&+\frac{g_{aa}}{2}\int dxdy \ \hat{\Psi}^\dagger(\mathbf{r})\hat{\Psi}^\dagger(\mathbf{r})\hat{\Psi}(\mathbf{r})\hat{\Psi}(\mathbf{r})~,
\end{align}
where $\mathbf{r} = (x,y)$ is a continuous coordinate.
Here, $\hat{\Psi}^\dagger$ ($\hat{\Psi}$) is the bosonic creation (annihilation) field operator for the atoms with mass $m$. The pump strength $\epsilon_p$ is in units of the recoil frequency $\omega_\mathrm{rec}=\hbar k_p^2/2m$ with the wavevector $k_p=2\pi/\lambda_p$. 
The short-range collisional interaction strength is parametrized by $g_{aa}$. 
We operate in the red-detuned regime where the pump frequency is negative detuned with respect to the atomic transition frequency, leading to a negative light shift per intracavity photon. 

The lattice potential is given by
\begin{align}
   V(\mathbf{r}) =& -\frac{V_0}{2}\left[\cos\left(\frac{k_{\mathrm{L}}}{\sqrt{2}}(x+y)\right) + \cos\left(\frac{k_{\mathrm{L}}}{\sqrt{2}}(x-y)\right)\right] \nonumber
   \\ &+ \frac{\Delta V}{4}\left[\cos\left(\frac{k_{\mathrm{L}}}{\sqrt{2}}x\right) + \cos\left(\frac{k_{\mathrm{L}}}{\sqrt{2}}y\right)\right]~,
\end{align}
where $V_0\geq0$ is the total lattice depth and $k_L=2 \pi/\lambda_L$ the wavevector of the lattice laser. 
This potential provides two inequivalent lattice sites in each unit cell, denoted by $A$ and $B$.
The relative potential depth $\Delta V= V_B - V_A$ between these sites can be adjusted so that it determines the relative occupation of the sites.
We assume that all sites in the $A$ sublattice consist entirely of $s$-orbitals and the $B$ sublattice contains $p_x$- and $p_y$-orbitals. 
The $s$-orbitals on $B$-sites are energetically far detuned and neglected.

Assuming a sufficiently deep optical lattice, we can expand the atomic field operator $\hat{\Psi}(x,y)$ into the basis of Wannier functions 
\begin{align}
    \hat{\Psi}(\mathbf{r})=&\sum_{\mathbf{r}_i\in B}\Big[ w_{p_x}(\mathbf{r}-\mathbf{r}_i) \hat{b}_{p_y,\mathbf{r}_i}+w_{p_y}(\mathbf{r}-\mathbf{r}_i) \hat{b}_{p_y,\mathbf{r}_i}\Big]\nonumber
    \\&+\sum_{\mathbf{r}_i\in A} w_s(\mathbf{r}-\mathbf{r}_i) \hat{b}_\mathrm{s,\mathbf{r}_i}~.
\end{align} 
where $\mathbf{r}_i = (x_i, y_i)$ are the discrete lattice positions, with $i$ labelling sites in the respective sublattice $A$ or $B$, e.g., $\mathbf{r}_i = a(n_i, m_i)$ with $a = \lambda_L/\sqrt{2}$ and $w_s(\mathbf{r}-\mathbf{r}_i)$, $w_{p_x}(\mathbf{r}-\mathbf{r}_i)$, $w_{p_y}(\mathbf{r}-\mathbf{r}_i)$ are the Wannier functions localized at site $\mathbf{r}_i$. 
For $\mathbf{r}$ located on an $A$-site, only the second term contributes. 
For $\mathbf{r}$ on a $B$-site, only the first term contributes.

\begin{figure}[t!]
	\centering
	\includegraphics[scale=1.05]{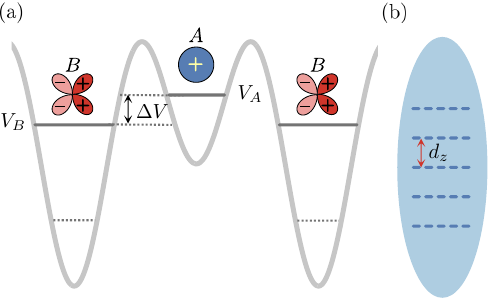}
    \caption{(a) Lattice potential along the $\hat{e}_x'$ direction with the inequivalent $A$ and $B$ sites. $A$-sites host $s$-orbitals with a potential depth $V_A$. $B$-sites host $p_x$ and $p_y$ orbitals at potential depth $V_B$. The potential difference is $V$. The p orbitals are sketched in red the s in blue.
    (b) Sketch of a single tube. Dashed lines indicate the artificial discretization with discretization length $d_z$. } 
    \label{fig:fig7}
\end{figure}

We further restrict our analysis to nearest neighbour tunneling processes, which in this lattice geometry corresponds to tunneling along the diagonal direction. 
The tunneling amplitude is 
\begin{align}
J= \int d (\mathbf{r}) w^{\nu *}(\mathbf{r}-\mathbf{r}_i) \left[-\frac{\hbar^2}{2m}\nabla^2+V(\mathbf{r})\right] w^{\nu'}(\mathbf{r'}-\mathbf{r'}_i)~.
\end{align}
The on-site interaction is 
\begin{align}
U_c= g_{aa}/2\int d \mathbf{r} w^{\nu_1 *} (\mathbf{r})w^{\nu_2 *} (\mathbf{r})w^{\nu_3 } (\mathbf{r})w^{\nu_3} (\mathbf{r})~.
\end{align}
Due to the odd parity of the $p$ orbitals, tunneling amplitudes depend on the bond orientation and acquire alternating signs. 
We neglect overall energy shifts and obtain the effective Hamiltonian in Eq.~\ref{eq:hamiltonian}.


\subsection{Ground state}
\label{app:sec-groundstate}

Here, we provide details on the origin of the chiral ground state and the emergence of staggered orbital angular momentum order in the $p$-band.
After the population transfer into the first excited band, the relevant degrees of freedom are the local $p_x$ and $p_y$ orbitals on each lattice site. The corresponding on-site interaction Hamiltonian reads
\begin{align}
\hat{H}_\mathrm{int}&=\frac{3U}{8}\left(\hat{n}_{p_x}(\hat{n}_{p_x}-1)+\hat{n}_{p_y}(\hat{n}_{p_y}-1)\right)+\frac{U}{2}\hat{n}_{p_x}\hat{n}_{p_y}\nonumber
\\&+\frac{U}{8}\left(\hat{b}^{\dagger }_{p_x}\hat{b}^{\dagger }_{p_x}\hat{b}_{p_y}\hat{b}_{p_y}+\hat{b}^{\dagger }_{p_y}\hat{b}^{\dagger }_{p_y}\hat{b}_{p_x}\hat{b}_{p_x}\right)~,
\end{align}
where the last term describes pair exchange between the two orbitals.
We assume 
\begin{align}
    b_{p_x}=\sqrt{n_x}, \quad b_{p_y}=\sqrt{ny}e^{i\phi}~.
\end{align}

Then the interaction energy depends on the relative phase as
\begin{align}
E(\phi)\propto \frac{U}{4} n_{p_x}n_{p_y} \cos(2\phi)~~.
\end{align}
This energy is minimized for
\begin{align}
\phi = \pm \frac{\pi}{2}~.
\end{align}
The corresponding single-particle state is
$\ket{p_\pm}=\frac{1}{\sqrt{2}}\left(\ket{p_x}\pm i \ket{p_y}\right)$, which carries a well-defined finite on-site angular momentum,
\begin{align}
\langle \hat{L}_z \rangle=\pm 1~.
\end{align}
In the lattice, tunneling between neighbouring sites introduces an additional staggered phase pattern due to the odd parity of $p$ orbitals.
Minimizing the total energy this leads to a staggered phase pattern in which neighbouring $B$-sites have opposite  angular momentum.
The global ground state has a staggered orbital angular momentum, spontaneous breaking of the $\mathbb{Z}_2$ symmetry and has vanishing total angular momentum, but finite staggered chiral order locally on the sites.


\subsection{Discretization of the tube direction}
\label{app:sec-tubes}

The lattice consists of one-dimensional tubes coupled in the $xy$-plane with a relatively shallow confinement along the $z$-direction, resulting in weakly bound atoms and pronounced atomic motion along the tubes.
We artificially discretizing the $z$-dimension by imposing an artificial lattice with spacing $d_z$, see Fig.~\ref{fig:fig7}~(b). 
The dynamics along the $z$-direction are described by a tight-binding model with tunneling amplitude $J_z$, yielding
\begin{align}
\hat{H}_z/\hbar =J_z \sum_{\mathbf{r},z}\Big(&\hat{b}^{\dagger}_{s,\mathbf{r},z}\hat{b}_{s,\mathbf{r},z+1}+\hat{b}^{\dagger}_{p_x,\mathbf{r},z}\hat{b}_{p_x,\mathbf{r},z+1}\nonumber
\\&+\hat{b}^{\dagger}_{p_y,\mathbf{r},z}\hat{b}_{p_y,\mathbf{r},z+1}+ \mathrm{h.c.}\Big).
\end{align}
The discretization length $d_z$ is chosen sufficiently small such that the associated kinetic energy dominates over all other energy scales. 
This ensures that the artificial lattice reproduces the continuum limit and does not impose additional physical constraints on the system.
We choose the tunneling amplitude on the order of the recoil frequency 
\begin{align}
    J_z=\omega_\mathrm{rec}~.
\end{align}
The discretization increases the number of lattice sites and thereby modifies the local density. The on-site interaction must be rescaled to preserve the correct interaction energy. 
We start from the three-dimensional contact interaction, we then integrate out the transverse degrees of freedom in $xy$ direction with the assumption of  harmonic confinement.
We obtain the effective one-dimensional interaction strength
\begin{align}
g_{1d} =\frac{2 \hbar^2 a_{\mathrm{s}}}{m x_0^2},
\end{align}
where $x_0$ denotes the transverse harmonic oscillator length and $a_{\mathrm{s}} = 5\,\mathrm{nm}$ is the three-dimensional scattering length.
This effective interaction strength describes a continuum model along the $z$-direction,
\begin{align}
H_{\mathrm{int}} =\frac{g_{1d}}{2}\int dz \,\hat{\Psi}^\dagger(z)\hat{\Psi}^\dagger(z)\hat{\Psi}(z)\hat{\Psi}(z).
\end{align}
When discretizing the $z$-direction, the integral is replaced by a sum such that the effective on-site interaction becomes
\begin{align}
U =\frac{g_{1d}}{d_z}.
\end{align}
This relation ensures that the interaction energy per unit length remains unchanged and that the discretized model correctly reproduces the continuum limit.


\section{Details Mean-field approximation}
\label{app-sec:meanfield}

We consider the Hamiltonian in Eq.~\ref{eq:hamiltonian} and employ the mean-field ansatz in the chiral basis 
\begin{align}
    \ket{\Phi_\mathrm{MF}}=\ket{\alpha}\otimes \prod_\mathbf{r}\ket{\beta_s}_\mathbf{r}\otimes\ket{\beta_\pm}_\mathbf{r}
\end{align}
The states $\ket{\beta_\pm}_\mathbf{r}$ are superpositions of the $p_x$ and $p_y$ orbitals at the same site with a $\pm \pi/2$ relative phase,
\begin{align}
&\ket{\beta_\pm}_{\mathbf{r}\in \mathrm{even}}\equiv\ket{\beta_+}_\mathbf{r} = \frac{1}{\sqrt{2}} \ket{\beta_{p_x} + i \beta_{p_y}}_\mathbf{r}~, 
\\&\ket{\beta_\pm}_{\mathbf{r}\in \mathrm{odd}}\equiv\ket{\beta_-}_\mathbf{r} = \frac{1}{\sqrt{2}} \ket{\beta_{p_x} - i \beta_{p_y}}_\mathbf{r}~.
\end{align}
For the $\Phi^+$ state, they act as follows
\begin{align}
    &\hat{\beta}_{+,\mathbf{r}}\ket{\beta_{+,\mathbf{r}}}=i(-1)^{r_y+0.5}\beta_{+}\ket{\beta_{+,\mathbf{r}}}
    \\& \hat{\beta}_{-,\mathbf{r}}\ket{\beta_{-,\mathbf{r}}}=i(-1)^{r_y-0.5}\beta_{-}\ket{\beta_{-,\mathbf{r}}}
    \\&\hat{\beta}_{s,\mathbf{r}}\ket{\beta_{s,\mathbf{r}}}=\theta(\mathbf{r})\beta_{s}\ket{\beta_{s,\mathbf{r}}}
\end{align}
with
\begin{align}
    \theta(\mathbf{r})=(-1)^{r_x}\Big[\frac{1+(-1)^{r_x+r_y}}{2}+i\frac{1-(-1)^{r_x+r_y}}{2}\Big]
\end{align}
Substituting these expressions gives the mean-field energy functional 
\begin{align}
E^\mathrm{MF}&(\alpha,\beta_s,\beta_+,\beta_-)
=- \Delta_\mathrm{c} |\alpha|^2 +  \frac{U}{2}\sum\limits_{j=1}^{M_s} \big( |\beta_s|^4-|\beta_s|^2\big)\nonumber\\
&
+ 2g\alpha_r \sum\limits_{j=1}^{M_p}\big(|\beta_+|^2-|\beta_-|^2\big)\nonumber\\ 
&
- 4 J \sum\limits_{j=1}^{M_p}\big( |\beta_s\beta_+|+  |\beta_s\beta_-|\big) \nonumber\\
&+\big(V_{B} + \Delta_{0}|\alpha|^{2}-\tfrac{3U}{8}\big) \sum\limits_{j=1}^{M_p}
\big(|\beta_+|^2+|\beta_-|^2\big)\nonumber\\
&
+ \frac{U}{2}\sum\limits_{j=1}^{M_p} \bigl( |\beta_+|^4+|\beta_-|^4\bigr).
\end{align}
The densities per site are defined as
\begin{equation}
\beta_s = \sqrt{\frac{N_s}{M_s}}, \quad
\beta_+ = \sqrt{\frac{N_e}{M_p}}, \quad
\beta_- = \sqrt{\frac{N_o}{M_p}},
\end{equation}
where $M_A$ and $M_B$ denote the number of lattice sites in the $A$- and $B$-sublattices, and $N_s, N_e, N_o$ are the total number of particles on each sublattice. Conservation of total particle number gives
\begin{equation}
N = N_s + N_e + N_o.
\end{equation}

We also define the total $p$-orbital occupation and the imbalance:
\begin{equation}
\bar{N}_p = N_e + N_o, \qquad
\Delta = N_e - N_o.
\end{equation}
Expressing the density per site in these terms,
\begin{equation}
    \begin{split}
        \beta_s & = \sqrt{\frac{N_s}{M_s}}, \ \beta_+  = \sqrt{ \frac{\bar{N}_p+\Delta}{2M_p}}, \ \beta_-  = \sqrt{\frac{\bar{N}_p-\Delta}{2M_p}}~,      
    \end{split}
\end{equation}
the energy then reads,
\begin{align}
    E^\mathrm{MF}&(\alpha,\Delta,\Bar{N}_p)
    =-\Delta_c |\alpha|^2 + \frac{U}{2 M_A}(N-N_p)^2 
    \\&+2g\alpha_r \Delta+N_p\big(V_B+\Delta_0|\alpha|^2-\frac{3U}{8}\big)\nonumber
    \\&-4J \sqrt{\frac{M_B(N-N_p)}{2M_A}} \Big( \sqrt{N_p+\Delta}+\sqrt{N_p-\Delta}\Big)\nonumber
    \\&+\frac{U}{4M_B}\big( \Delta^2+N_p^2\big)-\frac{U}{2}\big(N-N_p)~.\nonumber
\end{align}
Minimizing in terms of $\alpha$ with $\partial E/\partial\alpha=0$ gives
\begin{equation}
    \alpha(\Delta,\Bar{N}_p)= \frac{g \Delta}{-\Delta_c+\Delta_0\Bar{N}_p}~.
\end{equation}
To ensure identical boundary conditions, we consider a system with only $s$-orbitals at the boundary.
This gives the following relation for the total number of sites:
\begin{align}
    M_A=ZM^2, &&M_B=Z(M-1)^2~.
\end{align}
Leading to the final energy functional
\begin{align}
    E^\mathrm{MF}&(\Delta,\Bar{N}_p)
    =-\Delta_c \Big|\frac{g \Delta}{-\Delta_c+\Delta_0\Bar{N}_p}\Big|^2 + \frac{U}{2ZM^2}(N-N_p)^2 \nonumber
    \\&+2g\mathrm{Re}\Big( \frac{g \Delta}{-\Delta_c+\Delta_0\Bar{N}_p}\Big) \Delta+N_p\big(V_B+\Delta_0|\alpha|^2+\frac{U}{8}\big)\nonumber
    \\&-4J \sqrt{\frac{(M-1)^2(N-N_p)}{2M}} \Big( \sqrt{N_p+\Delta}+\sqrt{N_p-\Delta}\Big)\nonumber
    \\&+\frac{U}{4Z(M-1)^2}\big( \Delta^2+N_p^2\big)~.
\end{align}
Constant energy shifts are neglected.


\end{document}